\documentclass{article} 
\usepackage{iclr2026_conference,times}

\usepackage{amsmath,amsfonts,bm}

\def\eqref#1{equation~\ref{#1}}

\def\1{\bm{1}}

\DeclareMathAlphabet{\mathsfit}{\encodingdefault}{\sfdefault}{m}{sl}
\SetMathAlphabet{\mathsfit}{bold}{\encodingdefault}{\sfdefault}{bx}{n}

\usepackage{multirow}
\usepackage{hyperref}
\usepackage{url}
\usepackage{algorithm}
\usepackage{algpseudocode}
\usepackage{graphicx}
\usepackage{subcaption}
\usepackage{booktabs}

\title{Learning Where the Physics Is: Probabilistic Adaptive Sampling for Stiff PDEs}

\iclrfinalcopy
\author{Akshay Govind Srinivasan, Balaji Srinivasan\\
Indian Institute of Technology, Madras\\
Chennai, TN 600107, India \\
\texttt{me22b102@smail.iitm.ac.in, sbalaji@dsai.iitm.ac.in} \\
}

\begin{document}

\maketitle

\begin{abstract}
Modeling stiff partial differential equations (PDEs) with sharp gradients remains a significant challenge for scientific machine learning. While Physics-Informed Neural Networks (PINNs) struggle with spectral bias and slow training times, Physics-Informed Extreme Learning Machines (PIELMs) offer a rapid, closed-form linear solution but are fundamentally limited by physics-agnostic, random initialization. We introduce the Gaussian Mixture Model Adaptive PIELM (GMM-PIELM), a probabilistic framework that learns a probability density function representing the ``location of physics'' for adaptively sampling kernels of PIELMs. By employing a weighted EM algorithm, GMM-PIELM autonomously concentrates radial basis function centers in regions of high numerical error, such as shock fronts and boundary layers. This approach dynamically improves the conditioning of the hidden layer without the expensive gradient-based optimization(of PINNs) or Bayesian search. We evaluate our methodology on 1D singularly perturbed convection-diffusion equations with diffusion coefficients $\nu=10^{-4}$. Our method achieves $L_2$ errors up to $7$ orders of magnitude lower than baseline RBF-PIELMs, successfully resolving exponentially thin boundary layers while retaining the orders-of-magnitude speed advantage of the ELM architecture.
\end{abstract}

\section{Introduction}

Modeling stiff ODEs/PDEs with sharp gradients poses severe resolution challenges~\citep{Madhavi2025}. While high-fidelity classical methods exist, they are often cost-prohibitive due to stability constraints and complex meshing requirements~\citep{Yagawa2011}. This motivates the development of mesh-free, model-based solvers to efficiently resolve stiff dynamics. Physics-Informed Neural Networks (PINNs)~\citep{RAISSI2019686} offer a versatile mesh-free framework but suffer from slow training and hyperparameter sensitivity. Recent benchmarks indicate they often require significantly more wall-clock time than classical solvers to reach comparable accuracy, particularly for stiff PDEs~\citep{McGreivy2024, odil2024inverse}. Furthermore, spectral bias and optimization pathologies frequently lead to the under-resolution of sharp gradients, even in overparameterized networks~\citep{krishnapriyan2021characterizing}

Physics-Informed Extreme Learning Machines (PIELMs)~\citep{dwivedi2019physics_pielm} substitute iterative backpropagation with a closed-form linear least-squares solution, achieving orders-of-magnitude acceleration. Yet, the efficacy of PIELMs is fundamentally limited by their initialization: random, physics-agnostic hidden features often suffer from spectral mismatch with the underlying stiff dynamics, leading to ill-conditioned representations~\citep{huang2006extreme,EmergentMindPIELM2025}. Although Radial Basis Function variants like RBF-PIELMs~\citep{DWIVEDI2025130924, srinivasan2025deep} introduce interpretable, localized receptivity, they typically rely on static node allocations that lack the flexibility to track evolving discontinuities and require manually designed problem-specific heuristics. Consequently, there remains a critical need for principled, data-driven adaptation mechanisms that can dynamically align basis functions with the moving wavefronts of singular perturbation problems while retaining the computational efficiency of linear solvers.

To tackle these challenges, this work introduces the \textbf{Gaussian Mixture Model Adaptive PIELM (GMM-PIELM)}, a probabilistic training framework that treats the PDE residual field as an error density over the computational domain. We postulate that the $\log(1 + |\text{residual}|)$ field can be interpreted as unnormalized probability density function representing the ``location of physics''. We then learn this distribution using an Expectation--Maximisation (E-M) based algorithm. GMM-PIELM concentrates hidden-unit centers in regions of high numerical error. This adaptive redistribution improves the conditioning of the hidden-layer system and enhances expressivity exactly where the solution exhibits stiff behavior, while retaining the linear least-squares solve of the ELM architecture (unlike ~\cite{TANG2023111868}). Compared to KAPI-ELM ~\citep{EmergentMindPIELM2025}, that relies on iterative Bayesian optimization for center placement, our method relies on unsupervised learning of the probability density of the residual field. As demonstrated on 1D singularly perturbed convection--diffusion equations, the proposed method attains much lower $L_2$ errors than RBF-PIELM while capturing the complex boundary layer.

\paragraph{Contributions}This paper presents a EM-based algorithm to adaptively sample kernels for PIELMs. Specifically:
\begin{itemize}
    \item We present a fast and interpretable residual field based expectation maximization algorithm to adaptively sample kernels for solving stiff PDEs using RBF-PIELMs
    \item We evaluate the capability and efficiency of the  algorithm against RBF-PIELMs using a 1D boundary layer problem with a single and double boundary layer as benchmark.
    
\end{itemize}

\section{Mathematical Formulation}
We consider a stationary PDE defined on a bounded domain $\Omega \subset \mathbb{R}^d$ with boundary $\partial \Omega$:
\begin{equation}
\mathcal{L}[u](x) = f(x), \quad x \in \Omega
\quad
\mathcal{B}[u](x) = g(x), \quad x \in \partial \Omega
\end{equation}
where $\mathcal{L}$ is a linear differential operator, $\mathcal{B}$ is a boundary operator, $f(x)$ is a source term, and $g(x)$ is the boundary data.  In RBF-PIELM, We approximate the solution $u(x)$ using a Single-Hidden-Layer Feedforward Network (SLFN) with $N$ hidden neurons:
\begin{equation}
\hat{u}(x; \mathbf{\beta}) = \sum_{j=1}^{N} \beta_j \phi_j(x) = \sum_{j=1}^{N} \beta_j \exp\left(-{\|x - x_0^j\|^2 \over 2s_j^2} \right)
\end{equation}
where $\beta_j$ are the hidden unit weights and $\phi_j(x)$ are non-linear activations. $x^j_0$ and $s_j^2$ are the parameters of RBF kernel. Sampling these points randomly poses challenge especially for stiff problems (refer Appendix Section~\ref{appendix:ConditionNumber}). ~\cite{DWIVEDI2025130924, srinivasan2025deep} use a manually designed heuristic to sample these points. But these require prior knowledge of underlying physics and is challenging to adapt in dynamic problems. 

\subsection{GMM-PIELM: Gaussian Mixture Model-Based Kernel Adaptation}
The accuracy of the approximation $\widehat{u}$ in the previous section is heavily contingent on the placement of the basis centers $x_0^j$ and the choice of widths $s_j$, especially for stiff problems. We address this by proposing a probabilistic framework that allocates resources proportional to the local physical complexity of the problem.

\paragraph{Probabilistic Formulation}
Let $\mathcal{R}(x; \theta) = \mathcal{L}[\hat{u}_\theta](x) - f(x)$ denote the PDE residual field for the current approximation $\hat{u}_\theta$. Standard physics-informed approaches minimize the global $L_2$ norm of this residual, effectively averaging errors across the domain. However, for singularly perturbed problems, this global metric is dominated by smooth regions, often washing out the high-frequency errors localized at shocks or boundary layers~\citep{krishnapriyan2021characterizing}.

To address this, we postulate that the $\log(1 + |\text{residual}|)$ field acts as an unnormalized probability density function (PDF) representing the spatial concentration of approximation error. We provide justification for the choice of $\log$-transform in Appendix Section~\ref{appendix:TransformJustification}. We define the \textit{Residual Energy Density}, $p_{\text{res}}(x)$, as:
\begin{equation}
    p_{\text{res}}(x) = \frac{\log(1 + |\mathcal{R}(x; \theta)|)}{Z}, \quad \text{where} \quad Z = \int_{\Omega} \log(1 + |\mathcal{R}(z; \theta)|) \, dz.
    \label{eq:residual_pdf}
\end{equation}
Intuitively, $p_{\text{res}}(x)$ maps the ``location of physics,'' highlighting regions where the  spectral bandwidth is insufficient to capture the underlying dynamics~\citep{nabian2019efficient,wu2023comprehensive}. 
\paragraph{GMM-PIELM} Under this hypothesis, the optimal allocation of hidden neurons corresponds to a distribution of basis centers $\{\mathbf{c}_j\}_{j=1}^N$ that maximizes the likelihood of sampling from this residual landscape. Hence we model this distribution as a mixture of gaussian
\begin{equation}
x_0^j \sim p(x; \Theta) = \sum_{k=1}^K \pi_k\, \mathcal{N}(x \mid \mu_k, \Sigma_k),
\end{equation}
where $\Theta = \{\pi_k, \mu_k, \Sigma_k\}_{k=1}^K$ are the mixing coefficients, means, and covariances, respectively. To fit this model to the evolving solution, we employ a weighted EM framework to maximize the residual-weighted log-likelihood. In the \textbf{E-step}, we evaluate the current approximation and compute the "responsibility" $q_{ik}$—the posterior probability that the $i$-th collocation point $x_i$ (with residual weight $w_i = \log(1 + |\mathcal{R}(x_i)|)$) belongs to the $k$-th Gaussian component:
\begin{equation}
q_{ik} = \frac{\pi_k\, \mathcal{N}(x_i \mid \mu_k, \Sigma_k)}{\sum_{\ell=1}^K \pi_\ell \, \mathcal{N}(x_i \mid \mu_\ell, \Sigma_\ell)}.
\end{equation}
In the subsequent \textbf{M-step}, we update the GMM parameters to concentrate the components in regions of high error. The updated means $\mu_k^*$ and covariances $\Sigma_k^*$ are computed as:
\begin{equation}
N_k = \sum_{i=1}^{N_{eval}} w_i q_{ik}, \quad \mu_k^* = \frac{1}{N_k} \sum_{i=1}^{N_{eval}} w_i q_{ik} x_i, \quad \Sigma_k^* = \frac{1}{N_k} \sum_{i=1}^{N_{eval}} w_i q_{ik} (x_i - \mu_k^*)(x_i - \mu_k^*)^\top
\label{eq:sigma_update}
\end{equation}
where $N_{eval}$ is the number of evaluation points on a dense grid. The derivation of the parameter updates is provided in Appendix Section~\ref{appendix:derivation}. Finally, we augment the adaptive sampling with a set of uniformly distributed centers to ensure global domain coverage and prevent basis depletion in low-residual regions. We then determine the local kernel width $s_j$ based on the $k$-nearest neighbor distance to ensure consistent overlap even in highly clustered arrangements. 

\section{Numerical Experiments}
We evaluate the capability of the GMM-PIELM framework against RBF-PIELM on the singularly perturbed 1D steady-state convection--diffusion equation with single and double boundary layers. It is  a canonical benchmark for validating numerical schemes designed for stiff dynamics~\citep{roos2008robust}. The underlying PDE is defined on the domain $x \in (0,1)$ as:
\begin{equation}
     -\nu u_{xx} + u_x = 0
    \label{eq:convection_diffusion}
\end{equation}
where $\nu > 0$ represents the diffusion coefficient. The exact analytical solution is given in ~\cite{EmergentMindPIELM2025}. We assess performance in the stiff regimes of $\nu = 10^{-4}$. This equation serves as a fundamental model for high-P\'eclet-number transport phenomena encountered in fluid mechanics, isolating the physics of boundary layer formation where advective forces dominate diffusive ones.

The primary computational challenge in this regime is the formation of an exponentially thin boundary layer of width $\delta \sim \mathcal{O}(\nu)$ near the outlet ($x=1$). Resolving this feature typically necessitates prohibitive mesh densities ($N \gg 1/\nu$) for uniform solvers, while learning-based methods like PINNs suffer from pronounced spectral bias, often failing to capture the high-frequency transition without specialized loss weighting~\citep{krishnapriyan2021characterizing, wang2021pinn}. Similarly, standard PIELMs with randomized initialization frequently lack sufficient basis support in this narrow region, leading to ill-conditioned linear systems and degraded accuracy (refer Appendix Section~\ref{appendix:ConditionNumber}). Refer Appendix Section~\ref{appendix:ImplementationDetails} for implementation details.

\begin{figure}[htbp]
    \centering
        \begin{subfigure}[b]{0.46\linewidth}
        \centering
        \includegraphics[width=\linewidth]{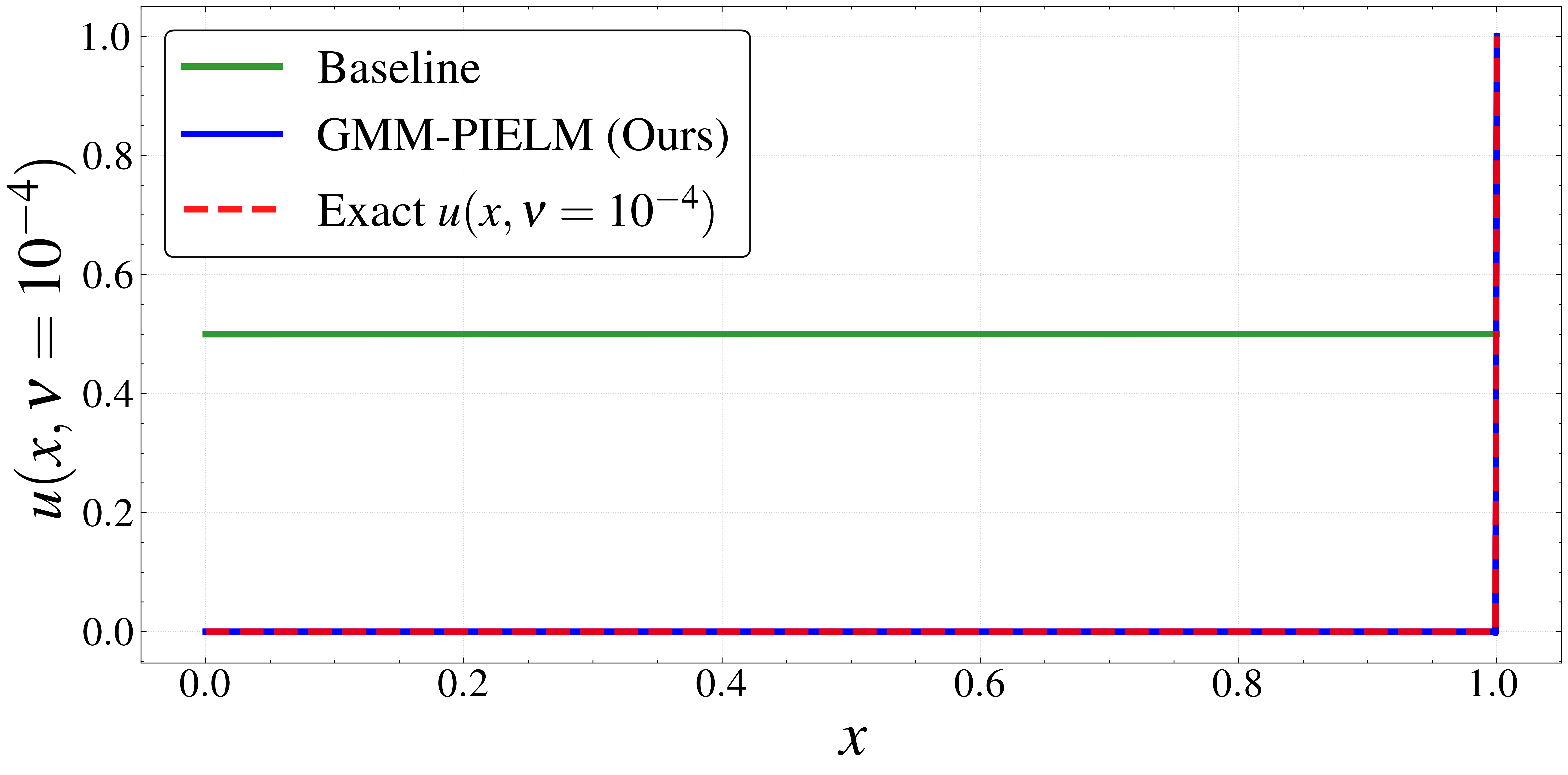}
        \caption{Solution profile for single boundary layer}
        \label{fig:fig_1_single_nu_1e4_1d}
    \end{subfigure}
    \hfill
    \begin{subfigure}[b]{0.46\linewidth}
        \centering
        \includegraphics[width=\linewidth]{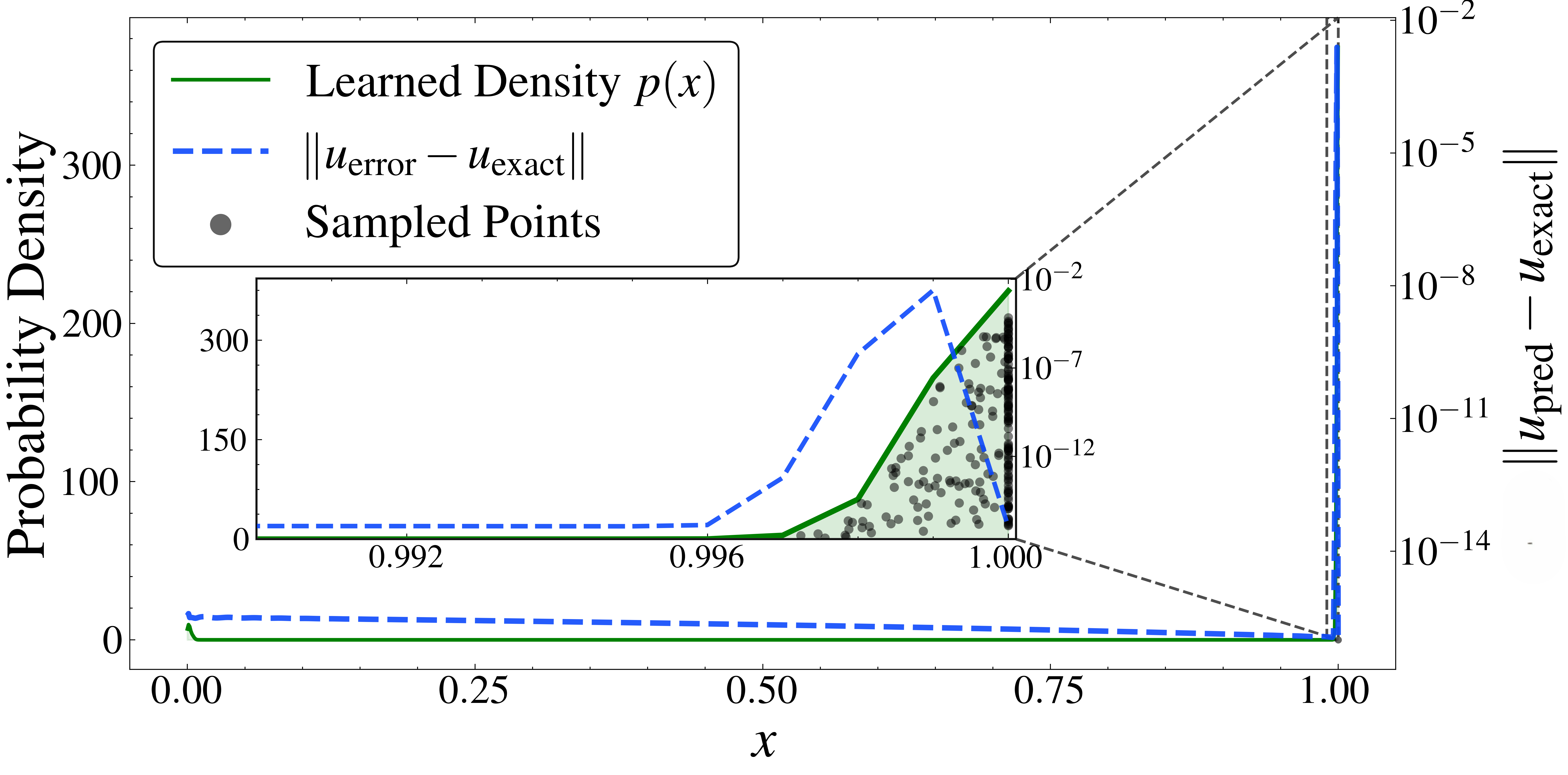}
        \caption{Error density and learned distribution for single boundary layer}
        \label{fig:fig_2_single_nu_1e4_1d}
    \end{subfigure}
    \begin{subfigure}[b]{0.48\linewidth}
        \centering
        \includegraphics[width=\linewidth]{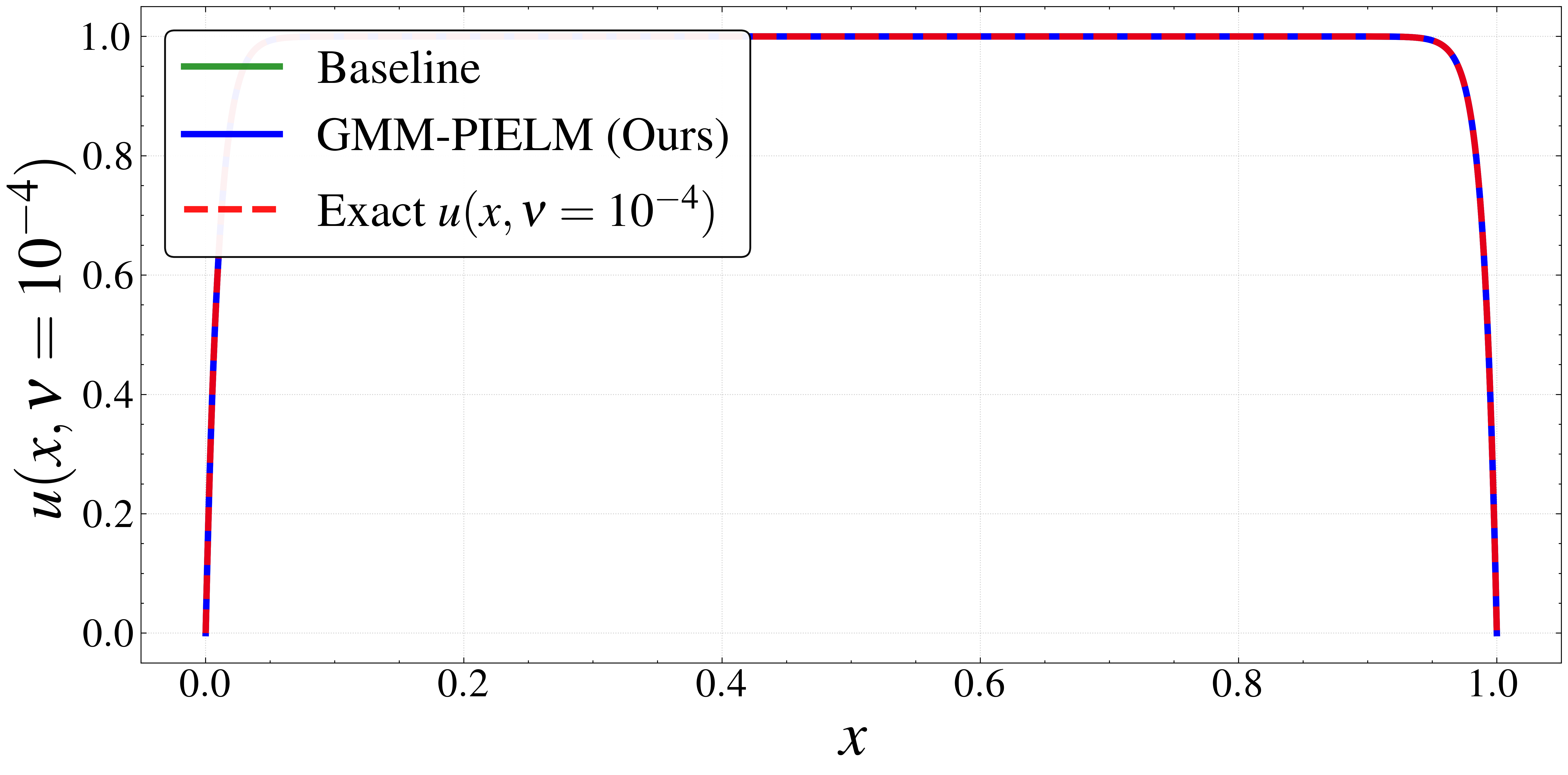}
        \caption{Solution profile for double boundary layer}
        \label{fig:fig_1_double_nu_1e4_1d}
    \end{subfigure}
    \hfill
    \begin{subfigure}[b]{0.48\linewidth}
        \centering
        \includegraphics[width=\linewidth]{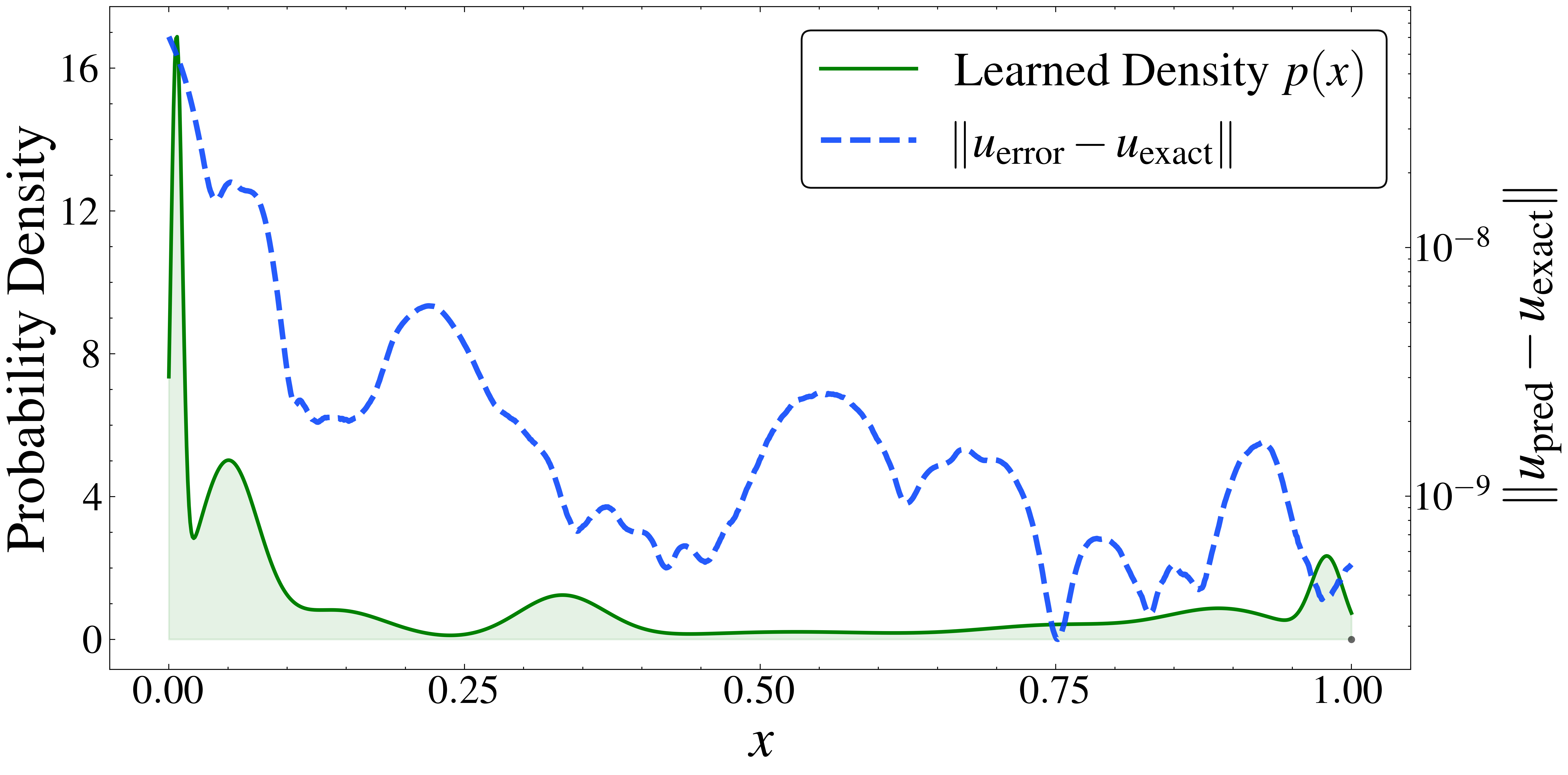}
        \caption{Error density and learned distribution for double boundary layer}
        \label{fig:fig_2_double_nu_1e4_1d}
    \end{subfigure}
    
    \caption{Performance analysis of GMM-PIELM on the 1D convection-diffusion equation with single and double boundary layer.}
    \label{fig:three_plots}
\end{figure}

\paragraph{Results} We observe from figures~\ref{fig:fig_1_single_nu_1e4_1d} and ~\ref{fig:fig_1_double_nu_1e4_1d} that our method (GMM-PIELM) is able to capture the sharp boundary layer at $x=1$ ($x=0,1$ in case of double boundary layer) as opposed to our baseline (RBF-PIELM). Our solution achieves upto $7$ orders of magnitude better accuracy as compared to RBF-PIELM (refer Table~\ref{tab:nu_comparison}). Additionally, we observe from figures~\ref{fig:fig_2_single_nu_1e4_1d} and~\ref{fig:fig_2_double_nu_1e4_1d} that the learned GMM distribution has similar structure to the error.

\begin{table}[t]
\centering
\begin{tabular}{l l l l}
\toprule
Boundary Condition & Method & Mean $L_2$ Error (RMSE) & Time (s)\\
\midrule
\multirow{2}{*}{Single ($N=300$)} 
& Baseline RBF-PIELM & $5.00 \times 10^{-1}$ & $0.071$ \\
& Ours (GMM-adaptive) & $2.73 \times 10^{-8}$ & $0.696$ \\
\midrule
\multirow{2}{*}{Double ($N=500$)} 
& Baseline RBF-PIELM & $1.01 \times 10^{-5}$ & $0.312$ \\
& Ours (GMM-adaptive) & $1.04 \times 10^{-9}$ & $1.776$ \\
\bottomrule
\end{tabular}
\caption{Comparison of Baseline vs. GMM-adaptive RBF-PIELM for single and double boundary layer problem ($\nu=10^{-4}$), showcasing significant improvement in accuracy.}
\label{tab:nu_comparison}
\end{table}

\vspace{-0.5em}

\section{Conclusion}
This work introduces the Gaussian Mixture Model Adaptive PIELM (GMM-PIELM), a framework that treats the PDE residual as a probability density representing the "location of physics." By interpreting the $\log(1 + |\text{residual}|)$ field as an unnormalized PDF, we use an EM algorithm to dynamically concentrate hidden-unit centers and widths in regions of high numerical error. This approach addresses the fundamental initialization limitations of standard Physics-Informed Extreme Learning Machines; instead of relying on physics-agnostic random sampling or heuristic node placement, the algorithm tracks sharp gradients in solution automatically. Numerical experiments on 1D singularly perturbed convection–diffusion equations demonstrate that GMM-PIELM achieves $L_2$ errors up to $7$ orders of magnitude lower than baseline RBF-PIELMs, successfully resolving thin boundary layers ($\nu=10^{-4}$) that standard methods fail to capture. While the adaptive process incurs a moderate computational overhead, it retains the fundamental speed advantages of the ELM architecture over gradient-based training. Consequently, GMM-PIELM offers a highly efficient and robust alternative for solving stiff, multi-scale physical systems that typically defy traditional mesh-less methods.
\paragraph{Future Work} Future Work will focus on extending this probabilistic framework to time-dependent PDEs, allowing the GMM centroids to track moving wavefronts in real-time. Additionally, we aim to investigate the scalability and stability of this approach to high-dimensional problems and complex geometries to enable it's use across multitude of applications.

\clearpage
\bibliography{iclr2026_conference}
\bibliographystyle{iclr2026_conference}

\appendix
\section{Appendix}
\subsection{The Condition Number Problem}
\label{appendix:ConditionNumber}
To determine $\mathbf{\beta} = [\beta_1, \dots, \beta_L]^T$, we substitute the ansatz $\hat{u}$ into the PDE and boundary conditions and evaluate them at a set of collocation points. Let $\{x_f^{(i)}\}_{i=1}^{N_f}$ be interior points and $\{x_b^{(k)}\}_{k=1}^{N_b}$ be boundary points.

For a linear operator $\mathcal{L}$, the residual constraints are linear in $\beta$:
\begin{equation}
\sum_{j=1}^{L} \beta_j \mathcal{L}[\phi_j](x_f^{(i)}) = f(x_f^{(i)})
\end{equation}
\begin{equation}
\sum_{j=1}^{L} \beta_j \mathcal{B}[\phi_j](x_b^{(k)}) = g(x_b^{(k)})
\end{equation}

This system can be written in matrix form as $\mathbf{H}\mathbf{\beta} = \mathbf{T}$, where:
\begin{equation}
\mathbf{H} = \begin{bmatrix}
\mathcal{L}[\phi_1](x_f^{(1)}) & \dots & \mathcal{L}[\phi_L](x_f^{(1)}) \\
\vdots & \ddots & \vdots \\
\mathcal{L}[\phi_1](x_f^{(N_f)}) & \dots & \mathcal{L}[\phi_L](x_f^{(N_f)}) \\
\lambda \mathcal{B}[\phi_1](x_b^{(1)}) & \dots & \lambda \mathcal{B}[\phi_L](x_b^{(1)}) \\
\vdots & \ddots & \vdots
\end{bmatrix}, \quad
\mathbf{T} = \begin{bmatrix}
f(x_f^{(1)}) \\
\vdots \\
f(x_f^{(N_f)}) \\
\lambda g(x_b^{(1)}) \\
\vdots
\end{bmatrix}
\end{equation}

Here, $\lambda$ is a penalty weight for boundary conditions. Since usually $N_f + N_b > L$, the system is overdetermined. The optimal weights in the least-squares sense are obtained via the Moore-Penrose pseudo-inverse:
\begin{equation}
\mathbf{\beta}^* = \mathbf{H}^\dagger \mathbf{T} = (\mathbf{H}^T \mathbf{H})^{-1} \mathbf{H}^T \mathbf{T}
\end{equation}

The accuracy of the least-squares solution depends heavily on the condition number $\kappa(\mathbf{H})$. In stiff PDEs, if the collocation points are uniformly distributed, the rows of $\mathbf{H}$ corresponding to the smooth regions are well-behaved, but the rows corresponding to the boundary layer (where gradients are huge) are sparse. If only a few points fall into the layer, the matrix $\mathbf{H}$ fails to capture the high-frequency dynamics, essentially treating the layer as noise. This results in a poor approximation $\beta^*$ that smooths over the shock. To fix this, we must increase the row density in the shock region, thereby explicitly constraining the solver to respect the physics in that zone.

\subsection{Algorithmic Implementation}
This probabilistic feedback loop is formalized in Algorithm \ref{alg:gmm_pielm}. The procedure alternates between a fast linear solve (ELM step) and a statistical adaptation of the basis functions (EM step) until the residual energy stabilizes.

\begin{algorithm}[H]
\caption{GMM-PIELM Adaptive Solving}
\label{alg:gmm_pielm}
\begin{algorithmic}[1]
\Require Operators $\mathcal{L}, \mathcal{B}$; Iterations $T$; Mixing Ratio $\alpha$
\State \textbf{Initialize:} Centers $x_0^{(0)}$ (uniform), Widths $s_j^{(0)}$ (constant)
\For{$t = 0$ to $T$}
    \State \textit{// Step 1: Linear Solve (PIELM)}
    \State Construct matrices $H^{\text{in}}, H^{\text{bd}}$ using current $\{x_0^j, s_j\}$
    \State Solve $\mathbf{w}^{(t)} = \arg\min_{\mathbf{w}} \| H \mathbf{w} - \mathbf{b} \|_2^2$
    \State Construct approx: $\widehat{u}^{(t)}(x) = \sum w_j \phi(x; x_0^j, s_j)$
    
    \State \textit{// Step 2: Assessment}
    \State Evaluate residual $\mathcal{R}(x; \theta) = \mathcal{L}[\widehat{u}^{(t)}] - f$ on grid
    \State Compute density $p_{\text{res}}(x) \propto \log(1 + |\mathcal{R}(x; \theta)|)$
    
    \If{$t < T$} 
        \State \textit{// Step 3: Adaptation (EM \& Hybrid Sampling)}
        \State Fit GMM parameters $\Theta$ to samples from $p_{\text{res}}(x)$
        \State Sample $x_{\text{gmm}} \sim p(x; \Theta)$ and $x_{\text{uni}} \sim \mathcal{U}(\Omega)$
        \State Update centers: $x_0^{(t+1)} \leftarrow \alpha x_{\text{gmm}} \cup (1-\alpha) x_{\text{uni}}$
        \State Update widths: $s_j^{(t+1)} \leftarrow \beta \cdot \text{dist}_k(x_0^j) + \epsilon$
    \EndIf
\EndFor
\State \textbf{Return} Final approximation $\widehat{u}^{(T)}$
\end{algorithmic}
\end{algorithm}
\subsection{Compute Specifications}
For Experiments, we run the experiment only on a CPU namely, a Ryzen 7 5800H with 16GB RAM. All processes are run on a single thread and process.
\subsection{Derivation of EM Steps}
\label{appendix:derivation}
The objective of the GMM-PIELM framework is to adaptively distribute the centers of the radial basis functions to match the spatial distribution of the PDE error. We postulate that the squared PDE residual field represents an unnormalized probability density function (PDF) indicating the ``location of physics.''

Let the \textbf{Residual Energy Density}, $p_{res}(x)$, be defined as:
\begin{equation}
    p_{\text{res}}(x) = \frac{\log(1 + |\mathcal{R}(x; \theta)|)}{Z}, \quad \text{where} \quad Z = \int_{\Omega} \log(1 + |\mathcal{R}(z; \theta)|) \, dz.
    \label{eq:residual_pdf}
\end{equation}
We approximate this target distribution using a Gaussian Mixture Model (GMM), $p(x; \Theta)$, parameterized by $\Theta = \{\pi_k, \mu_k, \Sigma_k\}_{k=1}^K$:
\begin{equation}
    p(x; \Theta) = \sum_{k=1}^{K} \pi_k \mathcal{N}(x | \mu_k, \Sigma_k)
\end{equation}

We seek the parameters $\Theta$ that make the model distribution $p(x; \Theta)$ best approximate the target residual distribution $p_{res}(x)$. The information-theoretic measure of the difference between two probability distributions is the \textbf{Kullback-Leibler (KL) Divergence}.

We minimize the KL divergence from the true distribution $p_{res}$ to the model $p_\Theta$:
\begin{align}
    \Theta^* &= \arg\min_{\Theta} D_{KL}(p_{res} || p_{\Theta}) \\
             &= \arg\min_{\Theta} \int_{\Omega} p_{res}(x) \ln \left( \frac{p_{res}(x)}{p(x; \Theta)} \right) dx \\
             &= \arg\min_{\Theta} \left[ \int_{\Omega} p_{res}(x) \ln p_{res}(x) dx - \int_{\Omega} p_{res}(x) \ln p(x; \Theta) dx \right]
\end{align}
The first term represents the entropy of the residual field. Since the residual $\mathcal{R}(x)$ is fixed during the adaptation step (it depends only on the previous iteration's weights), this term is constant with respect to $\Theta$. Minimizing the KL divergence is therefore equivalent to maximizing the second term (the cross-entropy):
\begin{equation}
    \Theta^* = \arg\max_{\Theta} \int_{\Omega} p_{res}(x) \ln p(x; \Theta) dx
\end{equation}
We approximate the integral using a dense set of collocation points $\{x_i\}_{i=1}^{N}$ via a Riemann sum. Substituting $p_{res}(x_i) \propto \log\left(1 + |\mathcal{R}(x_i)|\right)$:
\begin{equation}
    \Theta^* \approx \arg\max_{\Theta} \sum_{i=1}^{N} \log\left(1 + |\mathcal{R}(x_i)|\right) \ln p(x_i; \Theta)
\end{equation}
Defining the weight $w_i = \log(1 + |\mathcal{R}(x_i)|)$, we arrive at the \textbf{Weighted Log-Likelihood} objective function:
\begin{equation}
    \mathcal{L}(\Theta) = \sum_{i=1}^{N} w_i \ln \left( \sum_{k=1}^K \pi_k \mathcal{N}(x_i | \mu_k, \Sigma_k) \right)
\end{equation}

Direct maximization of $\mathcal{L}(\Theta)$ is intractable due to the summation inside the logarithm. We employ the EM algorithm. We introduce latent variables $z_{i} \in \{1, \dots, K\}$ indicating the component assignment for point $x_i$.

The expected complete-data log-likelihood (the Q-function) for the weighted dataset is:
\begin{equation}
    Q(\Theta, \Theta^{(t)}) = \sum_{i=1}^{N} w_i \sum_{k=1}^{K} q_{ik} \left[ \ln \pi_k + \ln \mathcal{N}(x_i | \mu_k, \Sigma_k) \right]
\end{equation}
where $q_{ik} = P(z_i = k | x_i, \Theta^{(t)})$ is the responsibility computed in the E-step (Eq. 5 in the paper).

We maximize $Q$ with respect to $\mu_k$ and $\Sigma_k$ by setting the gradients to zero.

Consider the terms in $Q$ dependent on $\mu_k$:
\begin{equation}
    J(\mu_k) = \sum_{i=1}^{N} w_i q_{ik} \left( -\frac{1}{2} (x_i - \mu_k)^T \Sigma_k^{-1} (x_i - \mu_k) \right)
\end{equation}
Taking the derivative with respect to $\mu_k$:
\begin{align}
    \frac{\partial J}{\partial \mu_k} &= \sum_{i=1}^{N} w_i q_{ik} \Sigma_k^{-1} (x_i - \mu_k)
\end{align}
Setting $\frac{\partial J}{\partial \mu_k} = 0$ and multiplying by $\Sigma_k$:
\begin{align}
    \sum_{i=1}^{N} w_i q_{ik} (x_i - \mu_k) &= 0 \\
    \sum_{i=1}^{N} w_i q_{ik} x_i &= \mu_k \sum_{i=1}^{N} w_i q_{ik}
\end{align}
Solving for $\mu_k$:
\begin{equation}
    \boxed{\mu_k^* = \frac{\sum_{i=1}^{N} w_i q_{ik} x_i}{\sum_{i=1}^{N} w_i q_{ik}}}
\end{equation}

Consider the terms dependent on $\Sigma_k$:
\begin{equation}
    J(\Sigma_k) = \sum_{i=1}^{N} w_i q_{ik} \left( -\frac{1}{2} \ln |\Sigma_k| -\frac{1}{2} (x_i - \mu_k)^T \Sigma_k^{-1} (x_i - \mu_k) \right)
\end{equation}
Using the matrix derivative identities $\frac{\partial \ln |\Sigma|}{\partial \Sigma} = \Sigma^{-1}$ and $\frac{\partial a^T \Sigma^{-1} a}{\partial \Sigma} = -\Sigma^{-1} a a^T \Sigma^{-1}$:
\begin{align}
    \frac{\partial J}{\partial \Sigma_k} &= \sum_{i=1}^{N} w_i q_{ik} \left( -\frac{1}{2} \Sigma_k^{-1} + \frac{1}{2} \Sigma_k^{-1} (x_i - \mu_k)(x_i - \mu_k)^T \Sigma_k^{-1} \right)
\end{align}
Setting to zero and multiplying by $\Sigma_k$ on both sides:
\begin{align}
    \sum_{i=1}^{N} w_i q_{ik} I &= \sum_{i=1}^{N} w_i q_{ik} (x_i - \mu_k)(x_i - \mu_k)^T \Sigma_k^{-1} \\
    \Sigma_k \left( \sum_{i=1}^{N} w_i q_{ik} \right) &= \sum_{i=1}^{N} w_i q_{ik} (x_i - \mu_k)(x_i - \mu_k)^T
\end{align}
Solving for $\Sigma_k$:
\begin{equation}
    \boxed{\Sigma_k^* = \frac{\sum_{i=1}^{N} w_i q_{ik} (x_i - \mu_k^*)(x_i - \mu_k^*)^T}{\sum_{i=1}^{N} w_i q_{ik}}}
\end{equation}

These equations exactly match Equation~\ref{eq:sigma_update} in the GMM-PIELM paper formulation.
\subsection{Justification on $\log$-transform}
\label{appendix:TransformJustification}
$p_{res}(x) \propto \log(1+|\mathcal{R}(x)|)$ is a critical mechanism for dynamic range compression, particularly in the stiff regimes targeted by our framework. Empirical tests using the raw squared residual field as the density concentrated the all the centers at the boundary, since the residuals in singularly perturbed problems vary by several orders of magnitude between the boundary layer and the rest of the domain. This led to a "collapse" of the basis distribution where global domain information was lost and the linear system became ill-conditioned in smoother regions. By applying the $\log$ transform, we successfully damp these extreme variations, allowing the Gaussian Mixture Model to resolve the "shoulders" of the error distribution rather than just the singularity. This ensures that while high-gradient regions are prioritized, sufficient information and basis support are maintained across the entire domain to yield a stable and accurate solution.
\subsection{Implementation Details}
\label{appendix:ImplementationDetails}
\paragraph{Experimental Setup}
All experiments were implemented using \texttt{scikit-learn} for GMM estimation and \texttt{scipy} for linear algebra operations, with a fixed random seed of 42. For both single and double boundary layer experiments, we sample a fixed set of $N_{\text{eval}} = 1500$ collocation points uniformly from the domain $\Omega = (0, 1)$ to construct the linear system. Boundary conditions are enforced via soft constraints by appending the boundary coordinates to the collocation set.

\paragraph{Baseline RBF-PIELM}
The baseline model initializes the hidden layer with $N$ neurons ($N=300$ for single, $N=500$ for double boundary layers). The centers are sampled uniformly:
\begin{equation}
    x_0^j \sim \text{Uniform}(0, 1), \quad j=1, \dots, N.
\end{equation}
The widths are set to a constant value $s_j = \frac{|\Omega|}{N} \times 2.5$, consistent with the overlap analogy presented in \cite{DWIVEDI2025130924}.

\paragraph{Proposed: Adaptive GMM-PIELM}
The adaptive method refines the network architecture over $T$ iterations. In each iteration, we compute the residual field $\mathcal{R}(x; \theta)$ on the dense grid and construct the residual energy density $p_{\text{res}}(x)$ as defined in Eq.~\eqref{eq:residual_pdf}. We then fit a GMM to data sampled from $p_{\text{res}}(x)$ and generate a new set of centers $\{x_0^j\}_{j=1}^N$ using a \textit{hybrid sampling strategy}:
\begin{itemize}
    \item \textbf{Residual-Guided ($100\alpha$\%):} Sampled from the fitted GMM components to target high-error regions (boundary layers).
    \item \textbf{Global Uniform ($100(1-\alpha)$\%):} Sampled uniformly from $\Omega$ to act as a safety net against basis depletion in smooth regions.
\end{itemize}
To accommodate the non-uniform clustering of centers, the local widths $s_j$ are adapted using $k$-Nearest Neighbors ($k$-NN):
\begin{equation}
    s_j = \beta \cdot \text{dist}_k(x_0^j) + \epsilon,
\end{equation}
where $\text{dist}_k$ is the Euclidean distance to the $k$-th neighbor ($k=2$) and $\beta$ is an overlap scaling factor.

\paragraph{Hyperparameters}
Table~\ref{tab:impl_params} details the specific parameters used for the Single Boundary Layer (BL) and Double BL experiments.

\begin{table}[h]
\centering
\caption{Hyperparameter settings for implementation.}
\label{tab:impl_params}
\begin{tabular}{lcc}
\toprule
\textbf{Parameter} & \textbf{Single BL} & \textbf{Double BL} \\ \midrule
Diffusivity ($\nu$) & $10^{-4}$ & $10^{-4}$ \\
Number of Neurons ($N$) & 300 & 500 \\
GMM Components ($K$) & 8 & 16 \\
Hybrid Ratio ($\alpha$) & 0.7 & 0.7 \\
Adaptation Iterations ($T$) & 3 & 3 \\
Sigma Scaling ($\beta$) & 1.1 & 1.5 \\ 
Random Seed & 42 & 42 \\\bottomrule
\end{tabular}
\end{table}
\end{document}